\documentclass{epl}

\title{Heat conductivity in the presence of a quantized degree of freedom}
\author{Jun-Wen Mao\inst{1,2}\thanks{E-mail:\email{jwmao@zimp.zju.edu.cn}} \and You-Quan Li\inst{1}}

\institute{
  \inst{1} Zhejiang Institute of Modern Physics, Zhejiang University, Hangzhou
310027, P.R. China.\\
  \inst{2} Department of Physics, Huzhou Teachers College, Huzhou 313000,
P.R. China. }


\begin{document}

\maketitle

\begin{abstract}
We propose a model with a quantized degree of freedom to study the
heat transport in quasi-one dimensional system. Our simulations
reveal three distinct temperature regimes. In particular, the
intermediate regime is characterized by heat conductivity with a
temperature exponent $\gamma$ much greater than 1/2 that was
generally found in systems with point-like particles. A dynamical
investigation indicates the occurrence of non-equipartition
behavior in this regime. Moreover, the corresponding Poincar\'e
section also shows remarkably characteristic patterns, completely
different from the cases of point-like particles.
\end{abstract}

The problem of heat conduction in low-dimensional systems has been
an interesting subject for a long time due to the potential
applications of nanostructured materials and the importance in
understanding the physics of low-dimensional systems~(see review
~\cite{Lepri_1} and the references therein). A long standing issue
is the exploration of the necessary condition for a dynamical
system to obey the Fourier
law~\cite{Casati,Kaburaki,Lepri_2,Lepri_3,Narayan}. For this
purpose, a class of ``Billiard gas channel'' has been proposed
recently~\cite{Alonso_1,Li_1,Li_2,Alonso_2} to investigate the
connection between macroscopic heat transport and microscopic
dynamics. In these models, usually a series of scatterers are
placed periodically along two parallel walls, and point-like
particles without interaction frequently undergo specular
reflections on the boundary formed by the scatterers and the
walls. By tracing the motion of the particles, the energy
transport in a periodic structure is investigated. Despite
intensive simulations in recent years, the sufficient condition
for a dynamic system ensuring the Fourier law is still not clear.
For example, it was ever found that the system with positive
Lyapounov exponent~\cite{Alonso_1} obeys the Fourier law, while
the same result was obtained in some linear mixing systems with
zero Lyapounov exponent~\cite{Li_1,Li_2,Alonso_2}. Recently, a
modified Lorentz model where particles collide with fixed
freely-rotating scatterers was suggested to investigate the heat
transport in local thermal equilibrium~\cite{Mej¨ªa,Larralde}.
Normal transport is obtained in such system, indicating that the
transport coefficients are finite in the thermodynamic limit.

Although these simple models provide a framework in which heat
transport emerges, they assume point-like particles (heat
carriers), and fail to provide a clear understanding of
temperature effects on heat transport. In fact, along with the
dramatic achievements in nanotechnology, the control of heat
transport for nanoscopic devices becomes increasingly important.
Thus, it is crucial to understand the temperature dependence of
heat conductivity in low dimensional system. Since the billiard
model captures the underlying dynamics of the system, it will be
appropriate for us to study the heat conductivity from the
microscopic dynamic point of view.

In this paper we propose a simple model to study the heat
transport, which demonstrates different regimes in the heat
conductivity. We consider a quasi-one-dimensional billiard gas
channel in which the heat carriers possess an additional
(intrinsic) quantized degree of freedom (e.g., rotational
velocity) beyond two translational ones. In general, this kind of
degree of freedom can come from the composite nature of the
carriers, or from the confinement in a third dimension. We explore
the scattering of the particles with rough surfaces (modeled by
fixed half disks). Here we focus on inelastic events and take no
account of the energy or the momentum exchange between the
particles and the scatterers. Namely, a collision with the rough
surface involves the occurrence of energy transfer between the
translational and the rotational degrees of freedom. Our
simulations show that there are three regimes according to the
temperature exponent $\gamma$ in the heat conductivity $\kappa\sim
T^\gamma$. The transition from high temperature regime to
intermediate temperature regime occurs close to a characteristic
temperature $T_c$, below which the translational velocity
distribution deviates from the Boltzmann distribution while energy
is distributed unequally among the three degrees of freedom. In
addition, by plotting the Poincar\'e section we find the clear
emergence of strange attractor accompanying the frozen of rotation
degree of freedom at low temperatures.

\begin{figure} [h]\vspace{-8mm}
\onefigure{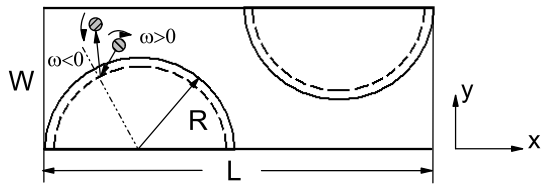} \vspace{-8mm}\caption{\label{fig:1}Schematic
of the collision of a particle of radius $r$ with a scatterer in a
unit cell of the channel. The surface of the semicircle scatterers
are plotted by dashed lines, near which the solid lines are the
surrounding tracks of the center of particles when colliding with
the scatterers. In the calculations, $L=3.6a$, $W=1.0392a$,
$R=0.89a$ and $r=0.01a$.}
\end{figure}

In our model, the channel in $x$ direction consists of two
parallel walls of distance $W$~(Figure~\ref{fig:1} shows the unit
cell of the channel). A series of semicircle scatterers with
radius $R$ are placed periodically along two horizontal walls.
Each cell has two semicircles, one on the bottom, the other on the
top. The heat baths at two terminals of the channel are modelled
by stochastic kernels of Gaussian type~\cite{Li_2},
\begin{eqnarray}
P(v_x)&=&\frac{m|v_x|}{k_B T}\exp(-\frac{m v_x^2}{2k_B T}),\nonumber\\
P(v_y)&=&\sqrt{\frac{m}{{2\pi k_B T}}}\exp(-\frac{m v_y^2}{2k_B
T}), \label{eq:Gaussian}
\end{eqnarray}
which determines the translational velocity $v_x$ and $v_y$ of the
particles in the channel after colliding with the walls between
the channel and the heat baths.

The particle with radius $r$ has a rotational velocity $\omega$ in
addition to the conventional velocities $v_x$ and $v_y$ which
describe the particle's translational movement. Since each unit
cell contributes only one heat carrier whose radius is small
compared with the scale of the channel, the interaction between
the particles is not taken into account for simplicity. Under the
assumption of $v_{\perp}'=-v_{\perp}$, the conservations of energy
and angular momentum read
\begin{eqnarray}
\frac{1}{2}mv_{\parallel}^{\prime2}+\frac{1}{2}I\omega^{\prime2}
 &=&\frac{1}{2}mv_{\parallel}^2+\frac{1}{2}I\omega^2,
\label{eq:energy} \\
mrv_{\parallel}^{\prime}+I\omega^\prime
&=&mrv_{\parallel}+I\omega, \label{eq:angular}
\end{eqnarray}
where the primed and unprimed ones represent the quantities after
and before a collision; the $v_{\parallel}$ and $v_{\perp}$ refer
to the components tangential and perpendicular to the scatterer
surface, respectively. The moment of inertia around an axis
through the center of a particle of mass $m$ is $I=\alpha m r^2$.
The similar collision rule has been appeared in
\cite{Mej¨ªa,Larralde}. Note that in our model $\alpha$ ranging
from $0$ to $1$ measures the moment of inertia, which corresponds
to the mass distribution of the disk particle, i.e., $\alpha=0$
for point particles, $1/2$ for uniform distributed mass.
Eqs.~(\ref{eq:energy}) and (\ref{eq:angular}) lead to the
following solution:

\begin{equation}\label{eq:omega}
\left( {\begin{array}{*{10}c}
  v_{\bot}^{\prime} \\[3mm]
  v_{\parallel}^{\prime} \\[3mm]
  \omega^\prime r \\
\end{array}} \right)  = \left( {\begin{array}{*{100}c}
   -1 & 0 & 0 \\[3mm]
   0 &\frac{1-\alpha}{1+\alpha} & \frac{2\alpha}{1+\alpha} \\[3mm]
   0 &\frac{2}{1+\alpha} & -\frac{1-\alpha}{1+\alpha}  \\
\end{array}} \right)  \left( {\begin{array}{*{100}c}
  v_{\bot} \\[3mm]
  v_{\parallel} \\[3mm]
  \omega r \\
\end{array}} \right).
\end{equation}

In addition to the specular reflections with the parallel walls,
the collision rules of semicircles are described as above
equation. Both the translational and the rotational velocities are
changed after a collision owing to the friction force against the
total tangential velocity of the contact point at the surface of
the particle (ball).

The rotational degree of freedom naturally introduces the
quantization condition $I\omega_l=l\hbar, \quad l=0,\pm 1,\pm 2,
\cdots$ which is well known for angular momentum in quantum
mechanics. Clearly, the moment of inertia determines the spacing
of the quantized angular velocity $\Delta
\omega=\hbar/I=\hbar/\alpha m r^2$, and the energy spacing
$\hbar^2/\alpha mr^2$ between the ground state and the first
excited state defines a characteristic temperature
$T_c=\hbar^2/\alpha mr^2 k_B$. Actually, the rotational part can
be regarded as an intrinsic degree of freedom, then our model
itself manifests internal excitations. If the energy for such
excitation is much less than $k_B T_c$, the internal degree of
freedom is frozen. In our simulations, the particle's
$\widetilde{\omega}$ takes an allowed discrete value near the
$\omega^\prime$ given by eq.~(\ref{eq:omega}). Under this
assumption, the conservation of angular momentum is broken down,
for the tangential component of velocity
$\widetilde{v_{\parallel}}$ is the same as
$v_{\parallel}^{\prime}$, while the perpendicular component
$\widetilde{v_{\perp}}$ is determined by the energy conservation.
In the simulation process, we set the parameter $a$, the particle
mass $m$, Boltzmann constant $k_B$ and Planck's constant $\hbar$
as unity for convenience.

As the quantized degree of freedom is assumed to play the role in
the channel, the rotational velocity then does not change when a
particle in the channel is rebounded from the walls of two heat
baths. The heat flux is calculated by the exchange of energy
carried by the particles via the collisions with a heat bath
within total time $t_M$ spent for $M$ such collisions \cite{Li_1},
\begin{equation}
J=\frac{1}{t_M}\sum_{j=1}^M (\Delta E)_j,\label{eq:heat flow}
\end{equation}
where $(\Delta E)_j=(E_{in}-E_{out})_j$ is the energy exchange at
the $j$th collision with a heat bath. Note that above heat flux is
for a single particle. In the channel with $N$ cells, the heat
conductivity is found to be $\kappa = J N^2 L / \Delta T$.

\begin{figure}[h]\vspace{-10mm}
\onefigure{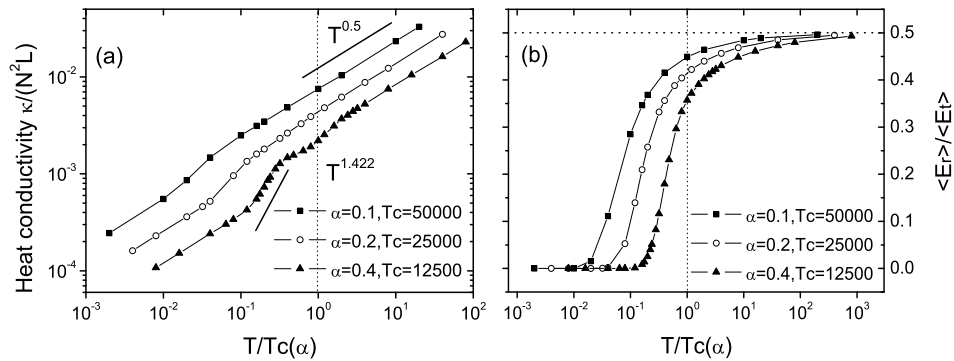} \vspace{-8mm}\caption{\label{fig:2} (a)
Temperature dependence of heat conductivity.  (b) The ratio of
average rotational energy over average translational energy at
different temperatures. The calculations are made for
$\alpha=0.1,0.2$, and $0.4$.}
\end{figure}

Figure~\ref{fig:2}(a) shows the heat conductivity-temperature
curves for a channel with $N=32$ cells. Here $T_c=5000/\alpha$ (in
units of $\hbar^2/m k_B a^2$). Because of non-equipartition of
energy in the system, which we will discuss next, one can not
define a local temperature from microscopic point of view. For
simplicity, we assume an average value of temperature
$T=(T_L+T_R)/2$ under small temperature gradient, where $T_L$ and
$T_R$ are the temperatures of two heat baths respectively. The
temperature difference $\Delta T/T$ is fixed at 0.2. To obtain the
heat conductivity correctly, we investigated heat flow $J$ in
Eq.(\ref{eq:heat flow}) as a function of temperature gradient in a
wide range of $5$ orders of magnitude, as was done in
ref.~\cite{Aoki}. We have verified the linear relation of
$J\sim\nabla T$ for temperatures and gradient explored in
fig.~\ref{fig:2}.  One can clearly see that the curves have larger
slope at temperatures near and less than $T_c$. Our calculations
exhibit that the maximum value of the slope reaches $1.422$ for
the case of $\alpha=0.4$.

To illustrate this characteristic, we investigate the average
translational energy $\langle E_t\rangle_{i}$
 and the rotational energy $\langle
E_r\rangle_{i}$ in a unit cell
\begin{equation}
\langle E_\sigma\rangle_{i}=\frac{\sum_{j=1}^M
t_{ij}(E_\sigma)_{ij}}{\sum_{j=1}^M t_{ij}}, \quad \sigma=t,\,r.
\end{equation}
where $E_t=\frac{1}{2}mv^2$ and $E_r=\frac{1}{2}I\omega^2$, $i$
represents the $i$th cell and $t_{ij}$ is the time spent within
the cell for the $j$th visit. In fig.~\ref{fig:2}(b) we plot the
ratio of $\langle E_t\rangle/\langle E_r\rangle$, where $\langle
E_t\rangle$ and $\langle E_r\rangle$ are the translational and the
rotational energies averaging over $N$ cells, respectively.
Clearly, there are three regimes with different characteristics.

1. Low temperature limit: Our simulations exhibit $\kappa$ $\sim$
$T^{\gamma}$ with $\gamma$ $\sim$ $1/2$ for various values of
$\alpha$ at low temperatures~(shown in fig.~\ref{fig:2}(a)). This
coincides with the result in previous literature~\cite{Li_1}. We
also calculate the temperature exponent $\gamma$ for the model
of~\cite{Alonso_1} and ~\cite{Mao}, respectively, obtaining the
result $\gamma\sim 1/2$ again. In the following we make an
estimation of the heat conductivity $\kappa$. In fact, at low
temperatures $T\ll T_c$, the heat carriers in our model can be
regarded as point-like particles because their thermal energy
$k_BT$ is much smaller than the first excitation energy. In this
case, the rotational motions are frozen. The heat conductivity is
estimated as $\kappa\sim\lambda\langle|\widetilde{v}|\rangle$
where $\lambda$ is the mean free path and
$\langle|\widetilde{v}|\rangle$ describes the average absolute
velocity~\cite{Gendelman}. For ideal gas it follows that
$\langle|\widetilde{v}|\rangle\sim \sqrt{2k_BT/m}$. Consequently,
we have $\kappa\sim T^{1/2}$ when $\lambda$ is treated as a
constant.

2. Intermediate temperature regime: The abrupt increase for the
curves in fig.~\ref{fig:2}(a) in the intermediate temperature
regime is related to the presence of quantized rotational degree
of freedom. It seems that the less moment of inertia the particle
possesses, the further the abrupt increasing regime is away from
its characteristic temperature. As is shown in
fig.~\ref{fig:2}(b), the ratio $\langle E_t\rangle/\langle
E_r\rangle$ rises rapidly from zero with increasing temperature
$T$ till $T_c$, then asymptotically to $1/2$. This indicates that
the total energy of the system is not equally partitioned among
the three degrees of freedom except in high temperature limit. As
a result, the temperature dependence of the average velocity
$\langle|\widetilde{v}|\rangle$ is significantly affected by the
ratio of the translational kinetic energy to the total energy,
which seems to be responsible for the abrupt increase of heat
conductivity since $\kappa$ is proportional to
$\langle|\widetilde{v}|\rangle$. The non-equipartition of energy,
which has been studied in granular
mixtures~\cite{Garzo,Feitosa,Dahl}, has now been observed in our
model due to non-conservation of angular momentum.

3. High temperature limit: In high temperature limit, the
saturation of the value of $\langle E_t\rangle/\langle E_r\rangle$
is observed. In this case, the quantum effect are almost
overwhelmed by the thermal fluctuations and thus the macroscopic
system is in thermodynamic equilibrium. The curves in
fig.~\ref{fig:2}(a) show $\gamma$ goes back to $1/2$ again,
however, the underlying mechanism is quite different.

\begin{figure}[h]\vspace{-10mm}
\onefigure{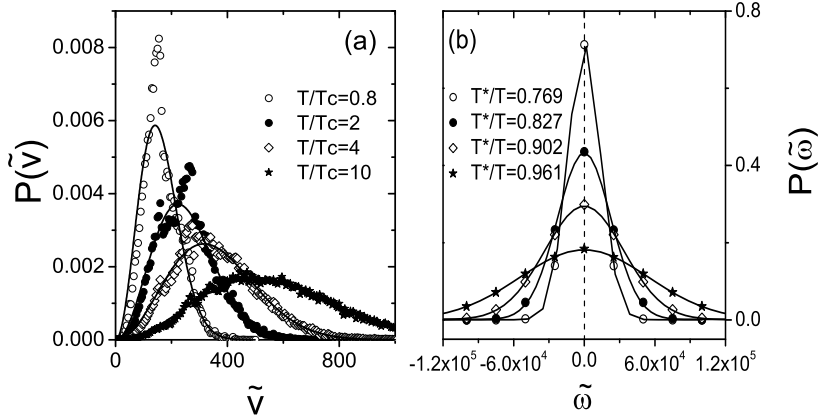} \vspace{-10mm}\caption{\label{fig:3}
Probability distribution at temperatures $T/T_c=0.8$ ($\circ$),
$2$ ($\bullet$), $4$ ($\diamond$), and 10 ($\star$) for
$\alpha=0.4$ and $N=8$. (a) The translational velocity
distribution. Solid lines are the Boltzmann distributions at the
corresponding temperatures. (b) The rotational velocity
distribution. The $T^*$ is obtained by the best fits~(solid lines)
of Gaussian distribution for numerical data.}
\end{figure}

In the non-equilibrium regime, the traditional definition of
temperature can not well describe the features of the system. To
understand the characteristics of these regimes further, we
measure the translational velocity distribution $P(\widetilde{v})$
and the rotational velocity distribution $P(\widetilde{\omega})$
of the particles. The simulation is performed in a equilibrium
system of $T_L=T_R$, where $10^5$ particles with initial Gaussian
distribution of velocity at a certain temperature $T$ are put into
the channel to make collisions with the semicircle rough surfaces,
while undergoing specular reflections with the straight walls.
After a long relaxation time, the system reaches a steady state.
As shown in fig.~\ref{fig:3}(a), solid lines are the Boltzmann
distributions in equilibrium at the corresponding temperatures.
Better agreements appear when $T/T_c\geq 10$, which indicates that
the particles can be considered as the classical ideal gas in
equilibrium in the high temperature limit despite the existence of
quantized rotational velocity. In comparison to the velocity
distribution at relatively low temperatures, the deviations from
the Boltzmann distribution are evident, leading to a larger
average velocity. Figure~\ref{fig:3}(b) shows the rotational
velocity distributions $P(\widetilde{\omega})$, which are typical
Gaussian distributions. $T^\ast$ is the best fits of the curves.
As expected, it is lower than the given temperature. The result of
$T^\ast/T\rightarrow 1$ with increasing temperature till the high
temperature limit shows that the system is asymptotically close to
equilibrium state.

\begin{figure}[h]\vspace{-5mm}
\onefigure{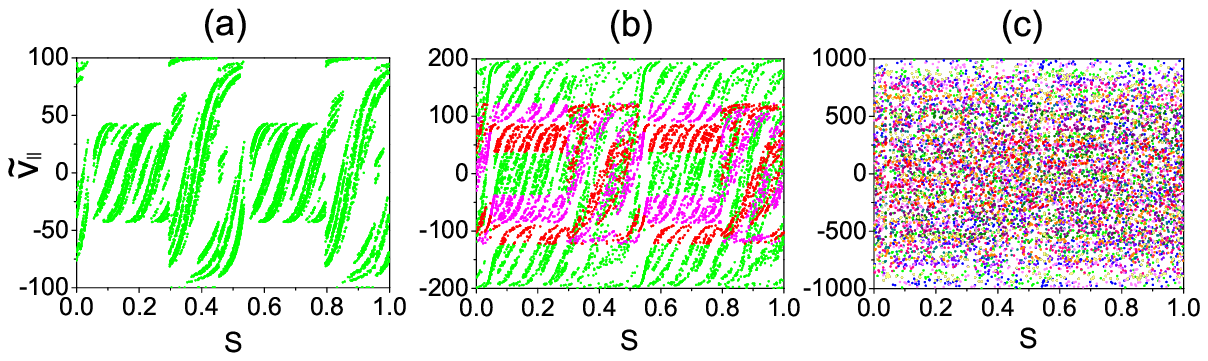}\vspace{-15mm}
\onefigure{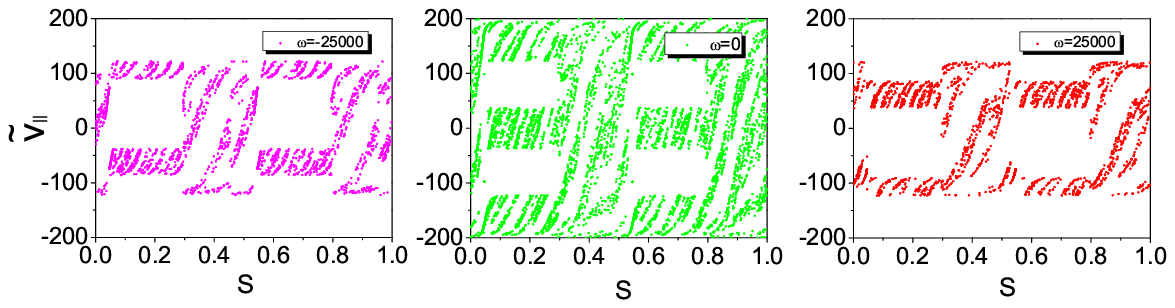}\vspace{-5mm}\caption{\label{fig:sos}
(Color online) Poincar\'{e} surface of section projected to
$s-\tilde{v_\|}$ plane, by following the first $10^4$ bounces with
zero initial angular velocity. The particle of velocity $v$ and
$\alpha=0.4$ starts from the center of the left boundary of the
cell with an incident angle of $0.9$. (a) $v=100$, and the allowed
$\tilde{\omega}=0$; (b) $v=200$, and the allowed
$\tilde{\omega}=-25000$, $0$, and $25000$; (c) $v=1000$, and more
values of $\tilde{\omega}$ are allowed. Additionally, the
Poincar\'{e} surfaces of section for $\tilde{\omega}=-25000$, $0$
and $25000$ at $v=200$ are displayed on the second row,
respectively.}
\end{figure}

To capture the physics of our model, we investigate the
Poincar\'{e} surface of section~(SOS) which has been employed to
characterize the microscopic dynamics for the motion of a
point-like particle colliding with the boundary~\cite{Berry}. It
is thus natural to investigate the dynamical characteristics of
the system in the presence of quantized degree of freedom. Using
our previous strategy~\cite{Mao}, we treat one unit cell of the
channel and close the two ends by straight frictionless walls.
Particles with certain initial velocity are injected into the
cell, then they collide with the boundary. The SOS in our model is
defined as ($s$, $\tilde{v_\|}$, $\tilde{\omega}$) where $s$ is
the distance to the starting site along the boundary,
$\tilde{v_{\|}}$ the tangential component of translational
velocities and $\tilde{\omega}$ the rotational velocity of the
particle. By projecting the points in phase space to the
$s-\tilde{v_{\|}}$ plane, we plot the Poincar\'{e} surface of
section ($s$, $\tilde{v_\|}$) where different values of rotational
velocity $\tilde{\omega}$ are displayed in different colors. In
our simulations, the initial incident angle brings about minor
effects on the SOS. Figure~\ref{fig:sos} (a), (b), and (c) show
three distinct cases of $v=100$, $200$, and $1000$, respectively.
As a comparison, we investigate the SOS for particles with a
classical rotational degree of freedom and find it is filled with
widely dispersed points. For the case of $v=100$ (shown in
figs.~\ref{fig:sos}(a)), the rotational velocity is frozen, thus
the points are confined in the plane $\tilde{\omega}=0$. Comparing
the pattern with that of classical situation, we note that many
areas are eclipsed. The remained pattern implies the existence of
a fractal attractor~\cite{Klages,Dellago}. This is completely in
agreement with what occurred in Bohr model of atoms. As is known
that the admissible orbits of an electron moving in the Coulomb
field of nucleus occupy the whole phase space. The Bohr-Sommerfeld
quantization condition excludes many of the classically admissible
orbits, which results in the emergence of eclipse in phase space.
When $v=200$, the rotational velocity is excited and the allowed
$\tilde{\omega}$ takes the values of $-25000$, $0$, and $25000$.
The corresponding SOS are shown in the bottom panels of
fig.~\ref{fig:sos}. The points are dispersed in the three layers
and form characteristic patterns in each layer. For the case of
$v=1000$, more excited states of higher rotational energy are
allowed, as shown in fig.~\ref{fig:sos}(c), and the points at
different $\tilde{\omega}$ are randomly dispersed.

In summary, we have proposed a method to introduce the quantized
degree of freedom for the first time in the study of heat
transport in the billiard channel. We have demonstrated three
characteristic regimes for heat conductivity in the presence of
quantized degree of freedom. Our results show that the
temperature-dependent interplay between the translational and
rotational degrees of freedom plays a crucial role in heat
conduction. We have also investigated the dynamical properties in
these regimes. The non-equipartition of energy between the
translational and rotational degrees of freedom is explicitly
shown except in the high temperature limit, and strange attractor
occurs remarkably in the low temperature limit. We expect that our
model provides a strategy to study the effects of the quantized
degree of freedom on heat transport.

\acknowledgments J.W. Mao thanks X. Wan for helpful discussion and
critical remarks on an earlier version of this paper. This work is
supported by NSFC No. 10225419 and HZNSF No. 2005YZ03.


\begin{thebibliography}{0}

\bibitem{Lepri_1}
  \Name{Lepri S., Livi R. \and Politi A.}
  \REVIEW{Phys. Rep.}{377}{2003}{1}.
\bibitem{Casati}
  \Name{Casati G., Ford J., Vivaldi F. \and Visscher W. M.}
  \REVIEW{Phys. Rev. Lett.}{52}{1984}{1861}.
\bibitem{Kaburaki}
  \Name{Kaburaki H. \and Machida M.}
  \REVIEW{Phys. Rev. Lett.}{181}{1993}{85}.
\bibitem{Lepri_2}
  \Name{Lepri S., Livi R. \and Politi A.}
  \REVIEW{Phys. Rev. Lett.}{78}{1997}{1896}.
\bibitem{Lepri_3}
  \Name{Lepri S., Livi R. \and Politi A.}
  \REVIEW{EuroPhys. Lett.}{43}{1998}{271}.
\bibitem{Narayan}
  \Name{Narayan O. \and Ramaswamy S.}
  \REVIEW{Phys. Rev. Lett.}{89}{2002}{200601}.
\bibitem{Alonso_1}
  \Name{Alonso D., Artuso R., Casati G. \and Guarneri I.}
  \REVIEW{Phys. Rev. Lett.}{82}{1999}{1859}.
\bibitem{Li_1}
  \Name{Li B., Wang L. \and Hu B.}
  \REVIEW{Phys. Rev. Lett.}{88}{2002}{223901}.
\bibitem{Li_2}
  \Name{Li B., Casati G. \and Wang J.}
  \REVIEW{Phys. Rev. E.}{67}{2003}{021204}.
\bibitem{Alonso_2}
  \Name{Alonso D., Ruiz A. \and Vega I. de}
  \REVIEW{Phys. Rev. E.}{66}{2002}{066131};
  \REVIEW{Physica D}{187}{2004}{184}.
\bibitem{Mej¨ªa}
  \Name{Mej¨ªa-Monasterio C., Larralde H., \and Leyvraz F.}
  \REVIEW{Phys. Rev. Lett.}{86}{2001}{5417};
\bibitem{Larralde}
  \Name{Larralde H., Leyvraz F. \and Mej¨ªa-Monasterio C.}
  \REVIEW{J. Stat. Phys.}{113}{2003}{197};
\bibitem{Aoki}
  \Name{Aoki K. \and Kusnezov D.}
  \REVIEW{Phys. Rev. Lett.}{86}{2001}{4029}.
\bibitem{Mao}
  \Name{Mao J. W., Li Y. Q. \and Ji Y. Y.}
  \REVIEW{Phys. Rev. E}{71}{2005}{061202}.
\bibitem{Gendelman}
  \Name{Gendelman O. V. \and Savin A. V.}
  \REVIEW{Phys. Rev. Lett.}{92}{2004}{074301}.
\bibitem{Garzo}
  \Name{Garzo V. \and Dufty J.}
  \REVIEW{Phys. Rev. E}{60}{1999}{5706}.
\bibitem{Feitosa}
  \Name{Feitosa K. \and Menon N.}
  \REVIEW{Phys. Rev. Lett.}{88}{2002}{198301}.
\bibitem{Dahl}
  \Name{Dahl S. R., Hrenya C. M., Garzo V. \and Dufty J. W.}
  \REVIEW{Phys. Rev. E}{66}{2002}{041301}.
\bibitem{Berry}
  \Name{Berry M. V.}
  \REVIEW{Euro. J. Phys.}{2}{1981}{91}.
\bibitem{Klages}
  \Name{Klages R., Rateitschak K. \and Nicolis G.}
  \REVIEW{Phys. Rev. Lett.}{84}{2000}{4268}.
\bibitem{Dellago}
  \Name{Dellago Ch., Glatz L. \and Posch H. A.}
  \REVIEW{Phys. Rev. E}{52}{1995}{4817}.
\end{thebibliography}
\end{document}